\documentclass[%
 aip,
 amsmath,amssymb,
reprint,%
]{revtex4-1}

\usepackage{graphicx}
\usepackage{dcolumn}
\usepackage{bm}
\usepackage{lipsum}

\usepackage[utf8]{inputenc}
\usepackage[T1]{fontenc}
\usepackage{mathptmx}
\usepackage{etoolbox}

\usepackage{graphicx}
\usepackage[dvipsnames]{xcolor}
\usepackage{appendix}
\usepackage{soul}

\makeatletter
\def\@email#1#2{%
 \endgroup
 \patchcmd{\titleblock@produce}
  {\frontmatter@RRAPformat}
  {\frontmatter@RRAPformat{\produce@RRAP{*#1\href{mailto:#2}{#2}}}\frontmatter@RRAPformat}
  {}{}
}%
\makeatother
\begin{document}

\preprint{AIP/123-QED}

\title[Magnetization patterns in GaAs-$\text{Fe}_{\text{33}}\text{Co}_{\text{67}}$ core-shell nanorods]{Magnetization patterns in GaAs-$\text{Fe}_{\text{33}}\text{Co}_{\text{67}}$ core-shell nanorods}

\author{Anastasiia Korniienko}
 \affiliation{Physics Department, Technical University of Munich, D-85748 Garching, Germany.}
\email{anastasiia.korniienko@tum.de.}

\author{Alexis Wartelle}%
\affiliation{Université Grenoble Alpes, CNRS, Institut Néel, Grenoble, 38042, France}

\author{Matthias Kronseder}
\affiliation{Institute for Experimental and Applied Physics, University of Regensburg, D93040 Regensburg, Germany}%

\author{Viola Zeller}
\affiliation{Institute for Experimental and Applied Physics, University of Regensburg, D93040 Regensburg, Germany}%

\author{Michael Foerster}
\affiliation{Alba Synchrotron Light Facility, CELLS, 08290 Cerdanyola del Vallès, (Barcelona), Spain}%

\author{Miguel Ángel Niño}
\affiliation{Alba Synchrotron Light Facility, CELLS, 08290 Cerdanyola del Vallès, (Barcelona), Spain}%

\author{Sandra Ruiz-Gomes}
\affiliation{Max Planck Institute for Chemical Physics of Solids, 01187 Dresden, Germany}%

\author{Muhammad Waqas Khaliq}
\affiliation{Alba Synchrotron Light Facility, CELLS, 08290 Cerdanyola del Vallès, (Barcelona), Spain}%
\affiliation{ Department of Condensed Matter Physics, University of Barcelona, 08028 Barcelona, Spain}%

\author{Markus Weigand}
\affiliation{Institut für Nanospektroskopie,Helmholtz-Zentrum Berlin für Materialien und Energie GmbH, 12489 Berlin, Germany}%

\author{Sebastian Wintz}
\affiliation{Institut für Nanospektroskopie,Helmholtz-Zentrum Berlin für Materialien und Energie GmbH, 12489 Berlin, Germany}%
\affiliation{Max Planck Institute for Intelligent Systems, 70569 Stuttgart, Germany}%

\author{Christian~H.~Back}
 \affiliation{Physics Department, Technical University of Munich, D-85748 Garching, Germany.}
 \affiliation{Center for Quantum Engineering (ZQE), Technical University Munich, D-85748 Garching, Germany.}

\date{\today}

\begin{abstract}
We present a study on the static magnetic properties of individual GaAs-$\text{Fe}_{\text{33}}\text{Co}_{\text{67}}$ core-shell nanorods. X-ray Magnetic Circular Dichroism combined with Photoemission Electron Microscopy (XMCD-PEEM) and Scanning Transmission X-ray Microscopy (STXM) were used to investigate the magnetic nanostructures. 
The magnetic layer is purposely designed to establish a magnetic easy axis neither along a high symmetry nor mirror axes to promote 3D magnetic helical order on the curved surface.
In practice, two types of magnetic textures
with in-plane magnetization were found inside the nanostructures' facets:
magnetic domains with almost longitudinal or almost perpendicular magnetization with respect to the axis of the tube.
We observe that a magnetic field applied perpendicular to the long axis of the nanostructure can add an azimuthal component of the magnetization to the previously almost longitudinal magnetization.

\end{abstract}

\maketitle
Recent progress in fabrication techniques unlocked a novel class of complex 3D magnetic structures with intriguing properties and effects not observable in traditional bulk or thin film materials~\cite{Streubel2016, Fernández-Pacheco_17}. There is a wide variety of techniques for producing nanostructures with complex geometries such as magnetic nanospirals~\cite{Fernandez-Pacheco2013}, Möbius strips~\cite{Skoric2020}, cylindrical nanowires with modulated diameter~\cite{Rodriguez2016}, rolled-up membranes~\cite{Streubel2012} and nanotubes with hexagonal cross-section~\cite{Ruffer}. Furthermore, interface engineering offers additional possibilities to control the magnetic behavior of the systems, e.g. by tuning overall anisotropies via combining magneto-crystalline anisotropies with those of interfacial origin.

A recent theoretical study~\cite{Sheka15} reveals that the geometry of the system may lead to the appearance of additional effective contributions to the magnetic energy landscape: an effective geometry-driven Dzyaloshinskii-Moriya interaction and an additional effective anisotropy.
This interplay of structure and magnetic properties enables the appearance of new complex magnetic configurations and also modifies the dynamics of magnetic excitations~\cite{Streubel2016, Sheka_2020a, Sheka_2022}. Curvature is the key parameter in this case; it manifests through symmetry-breaking effects, such as an asymmetry of the dispersion relation of spin waves in cylindrical nanotubes~\cite{Otalora_2016a, Otalora_2017a} and in nanotubes with hexagonal cross-sections\cite{Koerber_2021}. The considered structures~\cite{Otalora_2016a, Otalora_2017a, Koerber_2021} had an azimuthal magnetic state, i.e. vortex-like state; therefore, the magnetization was coupled to the curvature. By contrast, the longitudinal state would result in the magnetization uncoupled to the curvature. It would be of great interest to go beyond the sole vortex-like state and study a system with distinct easy axes, so the magnetization could be switched between almost longitudinal and almost azimuthal distributions.
This would allow control of the curvature's effect on the spin wave dispersion relation.

In this study, we investigate the magnetic properties of core-shell nanostructures, consisting of a $\text{Ga}\text{As}$ core with hexagonal cross-section and a $\text{Fe}_{\text{x}}\text{Co}_{\text{1-x}}$ magnetic shell, where $x = (33\pm3)\%$, capped by a thin $\text{Au}$ layer. The choice of material is based on the well-known magnetic properties~\cite{Muermann2008} of thin films of the $\text{Fe}_{\text{34}}\text{Co}_{\text{66}}$ alloy deposited on the particular $\text{Ga}\text{As}$ $ \{\overline{1}10\}$ crystallographic family of planes 
which make up the side facets of the $\text{Ga}\text{As}$ nanorods.
This alloy shows a thickness-dependent spin-reorientation transition~\cite{Muermann2008} and exhibits a bi-axial magneto-crystalline anisotropy for a wide range of thicknesses between 27 MLs (5.4 nm) and 58 MLs (11.6 nm).
These two easy axes are equivalent and symmetric around the [001] direction. Therefore, through careful tuning of the thickness of the magnetic layer, the magnetic easy axes can be oriented almost along the long nanostructure's axis and almost perpendicular to it. Such a system may thus provide control over the coupling strength between magnetization and curvature by switching the magnetization between these two axes. 

\begin{figure*}[hbt!]
\centering
\includegraphics[width=\textwidth]{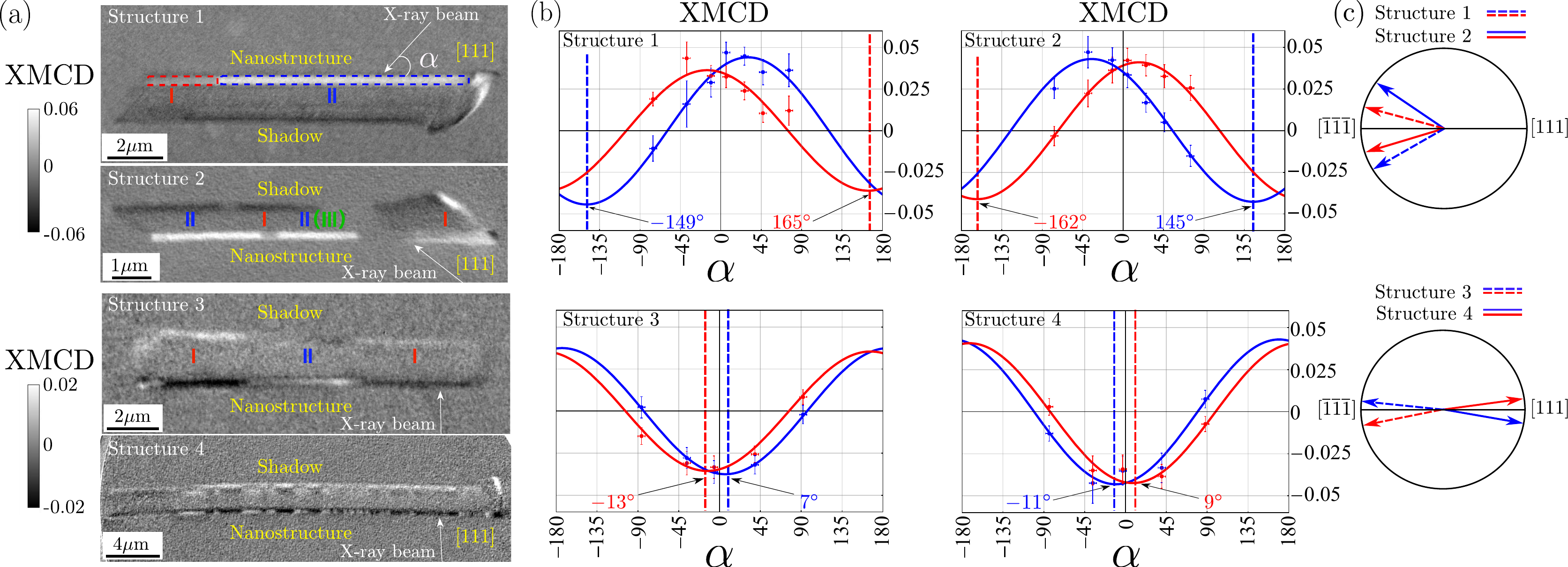}
\caption{ XMCD contrast: (a) XMCD contrast of the investigated nanostructures with schematics indicating the direction of propagation of the X-ray beam, the area with contrast from the structure and its shadow, and the different magnetic domains (marked by red and blue polygons). Nanostructures 1 and 2 were imaged before magnetic field application, and structures 3 and 4 were imaged after the application of a magnetic field of 42.5 mT perpendicular to the long axis of the structures and a subsequent magnetic field of 35.9 mT along the long axis of the structures. (b) Averaged angle-dependent XMCD contrast of all magnetic domains found in the nanostructures, 
and (c) schematics of the detected magnetization directions (in the plane of the facet). The blue and red dashed lines indicate the direction of magnetization, extracted from the fit. 
The data used to plot the angular-dependent XMCD contrast is shown in the supplementary.
Note that, due to the method of contrast calculation, the minima of XMCD contrast corresponds to the magnetization directed along with the beam.
}
\label{fig:XMCD-PEEM_angular-dependent-contrast}
\end{figure*}

The core-shell nanostructures are fabricated using Molecular Beam Epitaxy (MBE).  First, GaAs rods with hexagonal cross-section are grown on a $\text{GaAs}(111)$ wafer via the Vapor-Liquid-Solid mechanism in a III-V MBE chamber using Ga droplets as catalysts\cite{Wagner_64, Krogstrup_10}. Most of the nanostructures grow along the $[111]$ crystallographic direction (i.e. perpendicular to the surface of the wafer), however, oblique structures may occur with $[110]$ and $[001]$ growth directions. Note that only structures grown along the $[111]$ direction adopt hexagonal cross-sections with facets belonging to the $\{\overline{1}10\}$ crystallographic family.
After growth, the wafer is transferred in-situ to another MBE chamber in a pressure better than $1\times10^{-10}$ mbar, where 30 monolayers of $\text{Fe}_{\text{33}}\text{Co}_{\text{67}}$ (corresponding to 6.4 nm) 
are deposited at pressures of approximately $1\times10^{-10}$ mbar (base pressure $5\times10^{-11}$ mbar). The angle between the nanostructures' long axes and the Co and Fe evaporation direction is $28^\circ$. During evaporation, the wafer is rotated, producing a homogeneously thick layer of the $\text{Fe}\text{Co}$ alloy. The composition is monitored using X-ray Photoemission Spectroscopy (XPS), and the thickness calibration is given by Reflection High Energy Electron Diffraction measurements performed during deposition on a flat substrate. A $\text{Au}$ capping layer of 4 nm is then deposited to protect the structures from oxidation. The diameter of these nanostructures, defined as twice the apothem of the cross-section, varies between 400 nm and 700 nm. After growth, the wafer with the core-shell nanowires is shaken in isopropanol and treated with ultrasound. This detaches most (if not all) of the structures from their substrate, and they can be dispersed on a wafer suitable for experiments under an applied magnetic field of 2.3 T to align the structures.

\begin{figure*}[hbt!]
\centering
\includegraphics[width=\textwidth]{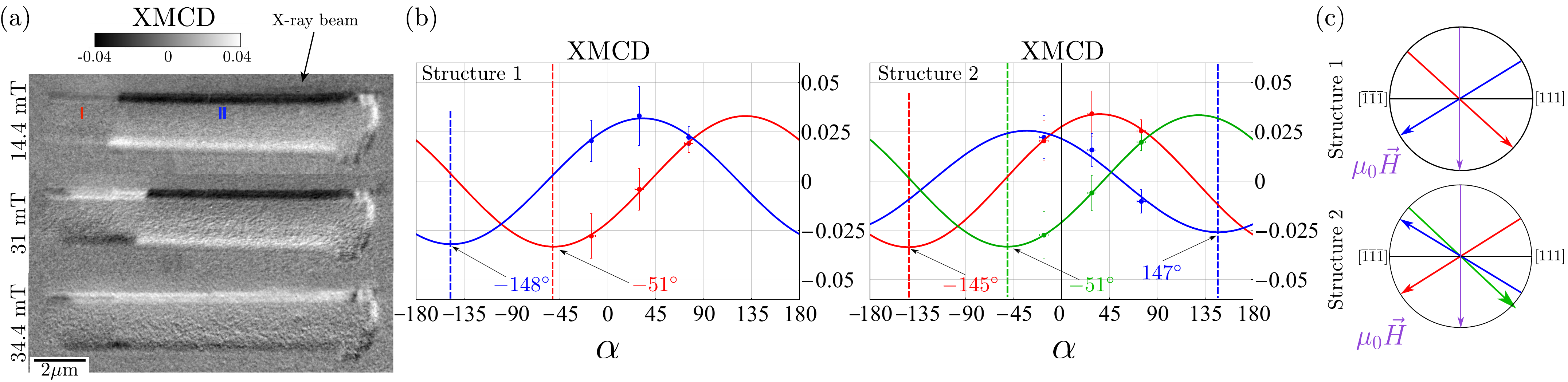}
\caption{ XMCD contrast: (a) The XMCD contrast for structure \#2 illustrates the magnetization switching from an almost longitudinal state to a state with a larger azimuthal component after the application of a magnetic field of 31 mT perpendicular to the structure. Imaging was done at remanence. (b) The angular-dependent XMCD contrast was acquired after the application of a magnetic field of 34.4 mT and (c) schematics of the extracted magnetization directions. The color coding corresponds to the one introduced above; the domain denoted in green switched from the central blue domain in the structure \#2. 
}
\label{fig:XMCD-PEEM_field-perpendicular}
\end{figure*}

Before describing the experimental results, let us first discuss the magnetic energy landscape of the presumably epitaxial ferromagnetic nanotubes.
The analysis of Ferromagnetic Resonance (FMR) measurements presented by Muermann \textit{et al.}~\cite{Muermann2008} on $\text{Fe}_{\text{34}}\text{Co}_{\text{66}}$ shows that a 30 ML thick FeCo alloy grown on GaAs (110) behaves as a biaxial system with easy axes $\pm 26^\circ$ with respect to the [001] direction, as a result of competition between a cubic magneto-crystalline anisotropy and an effective uniaxial anisotropy.
The interplay of these anisotropies would result in a magnetic state, known as magnetization easy cone~\cite{Coey} for a wide range of anisotropy constants $|K_1| \ge K_\mathrm{U}$ and $|K_1| \le 2|K_\mathrm{U}|$, where $K_1$ is the cubic anisotropy constant and $K_U$ is the uniaxial anisotropy constant.
In addition, the magnetostatics constrain the magnetization to lie in the plane of the FeCo film, and the equilibrium position would be defined by the ratio of anisotropy constants within the range specified above.
The uniaxial anisotropy favors the $[001]$ direction, while the minima of the cubic anisotropy lie along the $[-111]$ and $[1-11]$ directions in the case of a 32 ML thick $\text{Fe}_{\text{34}}\text{Co}_{\text{66}}$ film.
For the core-shell nanostructures with [111] growth direction, $ \{-110\}$ family of facets, and 32 ML thick FeCo shells, the easy axes would lie in the plane of the facets and assume directions at angles of $-29^\circ$ and $-81^\circ$ with respect to the growth direction of the GaAs nanowire.
They are equivalent and symmetric with respect to the [001] direction as a consequence of the uniaxial anisotropy pertaining to the [001] direction. The latter is found at $-\arccos(1/\sqrt{3}) \approx -55^\circ$ with respect to the [111] growth direction. However, deposition at an angle presumably slightly changes the system's anisotropies because of the self-shadowing effect~\cite{Alameda, Ozawa-Fe, Smith-Py}. In the case of similar core-shell nanostructures but with poly-crystalline $\text{Ni}_{80}\text{Fe}_{20}$ Permalloy (Py) magnetic shells, it was shown that deposition under such an angle leads to the appearance of a strong growth-induced uniaxial anisotropy leading, in the case of the Py tubes, to a magnetic vortex state~\cite{Zimmermann}.

To unravel the equilibrium magnetization directions and the domain configurations of the individual tubes, we performed magnetic imaging using X-ray Magnetic Circular Dichroism combined with Photoemission Electron Microscopy (XMCD-PEEM) at the CIRCE beamline of the ALBA Synchrotron Facility (Barcelona, Spain)\cite{Aballe2015}.
XMCD-PEEM is a powerful technique that can reveal the magnetic texture of an individual nanostructure, as was recently demonstrated by imaging of vortex and transverse magnetic domains in cylindrical nanowires\cite{Bran_2017a} and of the magnetic vortex configuration in magnetic tubes with hexagonal cross-section\cite{Wyss17}. 
The structures were illuminated with circularly polarized light at a grazing angle of $16^\circ$ with respect to the surface of the substrate, at the L$_3$ absorption edge of Co (778 eV). To reveal the magnetic configuration of the structures, a sequence of two PEEM images was taken with circular right (CR) and circular left (CL) polarization. The XMCD signal is then determined by calculating the asymmetry ratio $I_\text{XMCD} = \dfrac{I_\text{CL} -I_\text{CR} }{I_\text{CL} +I_\text{CR}}$ and the contrast is evaluated from the nanostructure, and not the shadow;  $I_\text{XMCD}$ will be referred to as XMCD in all figures.  The XMCD contrast is proportional to the projection of the magnetization onto the X-ray propagation vector~\cite{Jamet_15} $\bm k$,
 $I_{XMCD} \propto \bm m \cdot \bm k$. It should be noted that the data processing is done assuming that the magnetization lies completely in the plane of the facets of the nanostructures; this assumption is based on the hypothesis of dominating magnetostatics.
 With this in mind, we performed imaging of several nanostructures with several orientations of the X-ray beam with respect to the nanostructure’s long axis, but at a fixed inclination of $16^\circ$, see Supplementary for details. The angular dependence of the XMCD contrast reveals domains with magnetization direction neither along the long axes of the tubes nor perpendicular to them for all investigated structures, see Fig.~\ref{fig:XMCD-PEEM_angular-dependent-contrast}. Most domains exhibit small inhomogeneities in the XMCD contrast, which are accounted in the error bars of the measurement.

Qualitatively similar magnetization orientations were found for several nanotubes. In Fig.~\ref{fig:XMCD-PEEM_angular-dependent-contrast}(a), structure \#1 adopts two domains with tilt angles (defined with respect to the growth direction of GaAs core) $\alpha_1 = -149^\circ\pm 7^\circ$ and $\alpha_2 = 165^\circ\pm 10^\circ$, see Fig.~\ref{fig:XMCD-PEEM_angular-dependent-contrast}(b). Structure \#2 shows four magnetic domains of two types so that $\alpha_1 = -162^\circ\pm 5^\circ$ and $\alpha_2 = 145^\circ\pm 6^\circ$. The area without visible contrast is likely due to a damaged or absent $\text{Fe}_{33}\text{Co}_{67}$ layer. 
Interestingly, the other two investigated structures show symmetric behavior with large longitudinal magnetization components, even though one of the structures shows a large amount of short domains, while another only has a few long domains.
Structure \#3 shows domains with $\alpha_1 = -7^\circ\pm 3^\circ$ and $\alpha_2 = 13^\circ\pm 7^\circ$ and structure \#4 with $\alpha_1 = 9^\circ\pm 7^\circ$ and $\alpha_2 = -11^\circ\pm 8^\circ$. The extracted XMCD contrast from structures \#3 and \#4 suggests that the magnetization was rotated by $180^\circ$ after the application of a magnetic field of 16 mT along the nanostructure in $[\overline{1}\overline{1}\overline{1}]$ direction. In other words, the magnetization is tilted by the angles $\alpha_1$ and $\alpha_2$ counted from the $[\overline{1}\overline{1}\overline{1}]$ direction after switching.
The domain configurations of structures \#3 and \#4 were stable even after the application of a magnetic field of 31 mT along the nanotube's axis in $[\overline{1}\overline{1}\overline{1}]$ direction. The strength of the discussed magnetic field is of the order of anisotropy fields in the nanostructures, see appendix for the details.

\begin{figure*}[hbt!]
\centering
\includegraphics[width=0.7\textwidth]{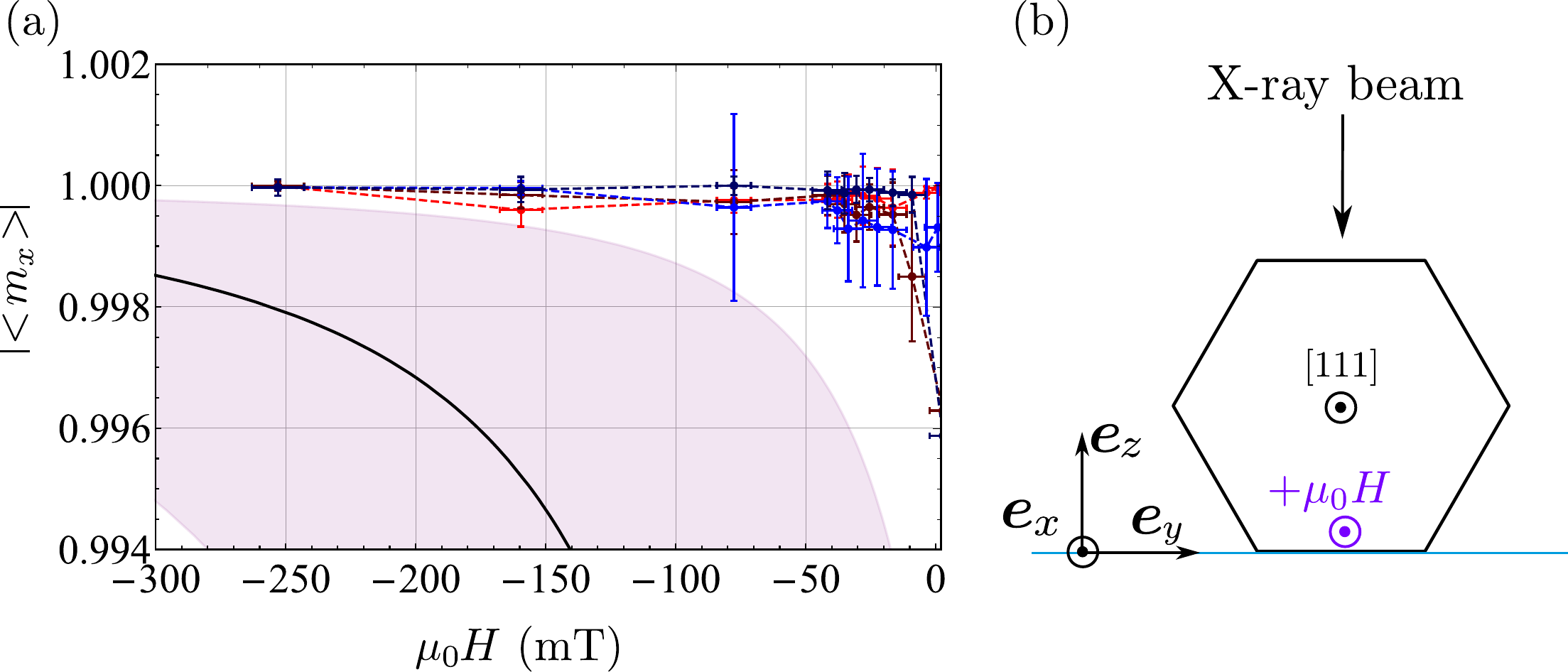}
\caption{(a) $\left| \left\langle m_x\right\rangle \right| $ and magnetization component evaluated for one pair of side facets (shown in red) and the opposite pair of side facets (shown in blue) of the core-shell nanostructure.  The bright (dark) shade of color indicates the direction of increasing (decreasing) strength of the magnetic field $H$ applied along the structure. The full black curve corresponds to the law of approach to saturation with $K_{1,\; M} \approx - 3.5 \cdot 10^4 \;  \text{J}/\text{m}^3$ and $K_{U,\; M} \approx 2.5 \cdot 10^4 \;  \text{J}/\text{m}^3$. The shaded area corresponds to the law of approach with constants in the range $( K_1, K_U) \in \left[\frac{1}{2}K_{1,\; M}, \; 2 K_{1,\; M}\right]  \times \left[\frac{1}{2}K_{U,\; M}, \; 2 K_{U,\; M}\right] $. With this, we demonstrate the incompatibility between our data and the constants reported by Muermann~\emph{et~al.}~\cite{Muermann2008}.
(b)~Schematics of STXM experiment.
}  
\label{fig:law_of_approach}
\end{figure*}

Images taken at remanence and after application of a magnetic field perpendicular to the structure \#1 reveal the appearance of a larger azimuthal component of the magnetization for the domain I, see Fig.~\ref{fig:XMCD-PEEM_field-perpendicular}. Moreover, the angular dependent XMCD contrast shows the appearance of an additional magnetization direction $-51^\circ$ for both structures \#1 (with an error of $5^\circ$ ) and \#2  (with an error of $3^\circ$ ), see Fig.~\ref{fig:XMCD-PEEM_field-perpendicular}. All fittings were performed considering the measurement error bars, and the uncertainties in the obtained angles were given by the mean deviations between the experimental data points and the continuous curves.

In contrast, structure \#4 features a large number of short magnetic domains even after the application of a magnetic field on the order of the anisotropy fields; this may result from the co-existence of the cubic zinc-blende (ZB) and hexagonal wurtzite (WZ) crystalline phases in the GaAs rods~\cite{Magri}. Indeed, the anisotropies in the $\text{Fe}_{33}\text{Co}_{67}$ shell depend on the crystallographic structure of the underlying GaAs facet, and different ZB and WZ crystalline domains will lead to different magnetic energy landscapes that may lead to different magnetic equilibrium states.

It has been shown~\cite{Stoner, Cullity} that ferromagnetic systems follow the law of approach to saturation when the magnetic field is not parallel to an easy or a hard magnetic direction and in the field range above irreversible processes in the system. Previously, the law of approach to saturation arising due to magneto-crystalline anisotropy in ferromagnets was considered in Ref.~\cite{ZHANG20102375}. Moreover, an approximate analytical solution valid in the field ranges larger than magnetocrystalline anisotropy fields was proposed, allowing to estimate the influence of the strength of anisotropy constants onto the law of approach to saturation. In order to experimentally determine the typical saturation fields, we measure hysteresis loops on an aligned array of oriented core-shell nanostructures (see appendix) with the magnetic field applied along the averaged nanostructures' long axis and perpendicular to it. 
The hysteresis loop in the geometry with magnetic fields parallel to the nanotubes axis shows that the magnetization mainly changes below 100 mT.
With this in mind, we perform Scanning Transmission X-ray Microscopy (STXM) measurements utilizing XMCD at the MAXYMUS end station at the BESSY II electron storage ring operated by the Helmholtz-Zentrum Berlin für Materialien and Energie, to measure a hysteresis loop on a single nanostructure, applying a magnetic field along the structure's long axis. As was mentioned above, the anisotropy landscape of $\text{Fe}_{\text{33}}\text{Co}_{\text{67}}$ would result in the $[111]$ direction being neither an easy nor a hard direction. The investigated nanostructure is first subjected to a magnetic field of +250 mT. Then, we measure the XMCD-STXM contrast under negative fields, see Supplementary for details.
Taking into account the geometry of the nanostructure, we convert the measured dichroic contrast into the (normalized) component of magnetization along the nanotube axis $m_x$, averaged over opposite pairs of side facets. The corresponding $\left| \left\langle m_x\right\rangle \right| $ is represented on Fig.~\ref{fig:law_of_approach}.
Following the approach reported in~\cite{ZHANG20102375} and using the anisotropy landscape for our $\text{Fe}_{33}\text{Co}_{67}$ shell, we determine the law of approach to saturation.
In the latter, we fix the value of the saturation magnetization to 1.98 T, to be consistent with independent FMR measurements on $\text{Fe}_{33}\text{Co}_{67}$~(110) films capped with aluminum. 
The law of approach to saturation determined with anisotropy constants ${K_{1,\;M} \approx - 3.5 \cdot 10^4 \;  \text{J}/\text{m}^3}$ and ${K_{U,\;M} \approx 2.5 \cdot 10^4 \;  \text{J}/\text{m}^3}$ reported by Muermann~\emph{et~al.}~\cite{Muermann2008}  does not allow to match the experimentally determined approach to saturation. Moreover, varying constants in the range ${( K_1, K_U) \in \left[\frac{1}{2}K_{1,\; M}, \; 2 K_{1,\; M}\right]  \times \left[\frac{1}{2}K_{U,\; M}, \; 2 K_{U,\; M}\right] }$ does not result in the approach to the magnetisation saturation revealed by measurements, see Fig.~\ref{fig:law_of_approach}. Smaller values of ${( K_1, K_U)}$ typically lead to faster saturation. As one can see, the measured magnetisation exhibits a faster evolution towards saturation, which may be caused by a growth-induced anisotropy which is not included in our model. Such effects can indeed be significant, as was discussed earlier in the case of Py~\cite{Zimmermann}. We point out that while a full Fe film shows a similar dependence of the growth-induced anisotropy on the incident angle as Py~\cite{Ozawa-Fe, Kambersky-Py}, the oblique-incidence anisotropy in Co films~\cite{Kamberska_Co} would favour longitudinal magnetization in the case of the nanostructure. Moreover, both our deposition angle ($62^\circ$ with respect to normal incidence on the facets) and the low shell thickness suggest~\cite{Kamberska_Co} contributions of the same order of magnitude as $K_{1,\;M}$, $K_{U,\;M}$.

In summary, we observed two types of magnetic textures in our 3D magnetic nanotubes namely with an easy axis either almost along or almost perpendicular to the nanotubes long axis.
The XMCD-PEEM technique doesn`t grant access to the magnetization of all the nanostructure`s facets and a 3D X-ray magnetic microscopy is required to reconstruct the complete 3D magnetic texture. Therefore, the observed configuration is a prerequisite for the formation of helical magnetic textures in 3D magnetic nanotubes.
We demonstrate that magnetic fields applied perpendicular to the nanotubes long axis introduce an azimuthal component of the magnetization in magnetic domains with initially almost longitudinal magnetization. 
On a more quantitative level, the insight we gained with STXM measurements points at magnetic anisotropies distinct from the full-film expectation.
The expected easy axes are $-29^\circ$ and $-81^\circ$ based on the values of the anisotropy constants reported by Muermann. However, our experimentally determined values do not align with these expectations.
Whether this discrepancy arises from the shell's deposition angle, thickness of magnetic layer, dispersion process or any other cause, it nevertheless does not suppress the targeted biaxial behaviour.

\begin{acknowledgements}
We acknowledge ﬁnancial support of the Deutsche Forschungsgemeinschaft through project ID 314695032 - SFB1277 subproject No. A08.
MF, MAN, MWH acknowlege funding by MICIN through project PID2021-122980OB-C54.
The XMCD-PEEM experiments were performed at CIRCE beamline at ALBA Synchrotron with the collaboration of ALBA staff.
We thank Helmholtz-Zentrum Berlin for the allocation of synchrotron radiation beamtime.
\end{acknowledgements}

\begin{appendices}
\section*{Appendix}

\begin{figure*}[hbt!]
\centering
\includegraphics[width=0.8\textwidth]{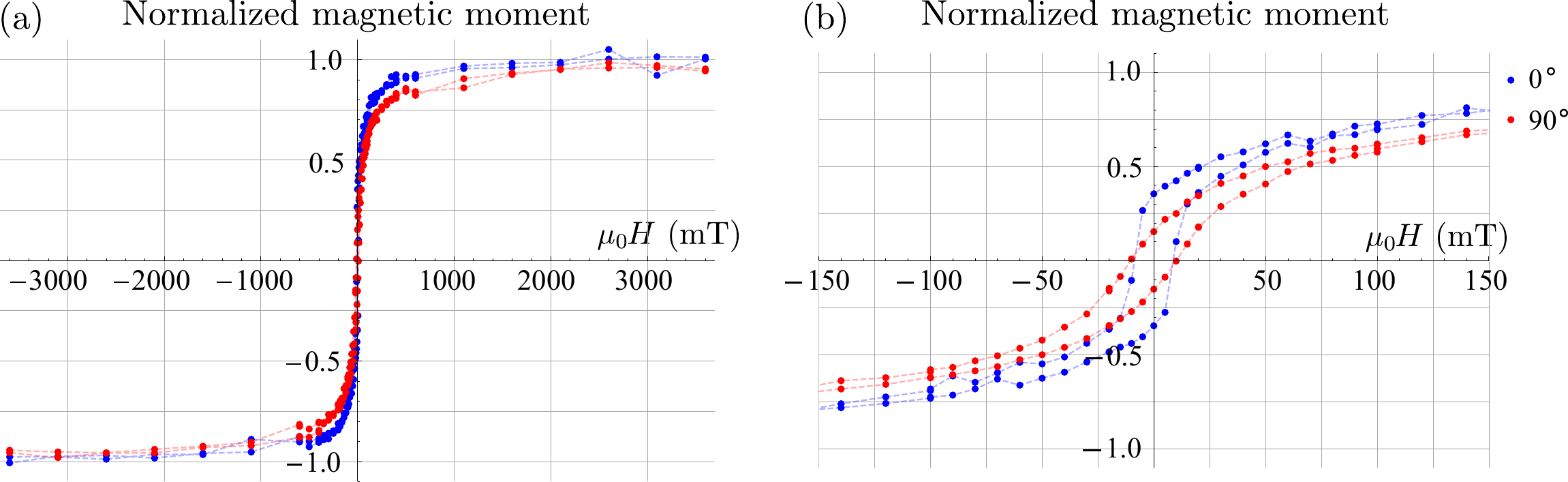}
\caption{
(a) Hysteresis loops for the array of aligned nanostructures with the field along the averaged long axis of nanostructures ($0^\circ$) and perpendicular to it ($90^\circ$) and (b) a zoomed-in view highlighting the coercive field and remanence.
}
\label{fig:SQUID}
\end{figure*}
Here, we report the hysteresis loops measured via SQUID (superconducting quantum interference device) on an aligned array of nanostructures. 
The structures were transferred from the wafer on which they were grown into isopropanol via ultrasonication. After that, they can be dispersed onto a substrate under a magnetic field of 2.3 T to align them.
The loop has a rather square-like shape when the magnetic field is applied along the average long axis of the nanostructures and reveals that about $80\%$ of change in magnetization happens below 150 mT (Fig.~\ref{fig:SQUID}). Moreover, the loop is closing in the field range below 100 mT, indicating that only reversible processes, \emph{i.e.} magnetization rotation, are expected to occur above 100 mT.
It should be noted that the hysteresis loops measured on the arrays of structures are affected by a certain spread in the direction of the long axis due to misalignment of individual structures.
The loops acquired with the magnetic field perpendicular to the structures hint that it is more difficult to magnetize the nanostructures in the direction perpendicular to the long axis since it requires pulling the magnetization out-of-plane of the side facets, see Fig.~\ref{fig:SQUID} (a) and (b). Note, that due to the spread in the direction of the long axis due to misalignment of individual structures and dipolar interactions between them, we do not apply the approach of saturation to the measurements on the array of nanostructures.

\end{appendices}
\nocite{*}
\bibliography{biblio}

\end{document}


\preprint{AIP/123-QED}

\title[Supplementary]{Supplementary}

\begin{figure*}
\centering

\includegraphics[width=0.65\textwidth]{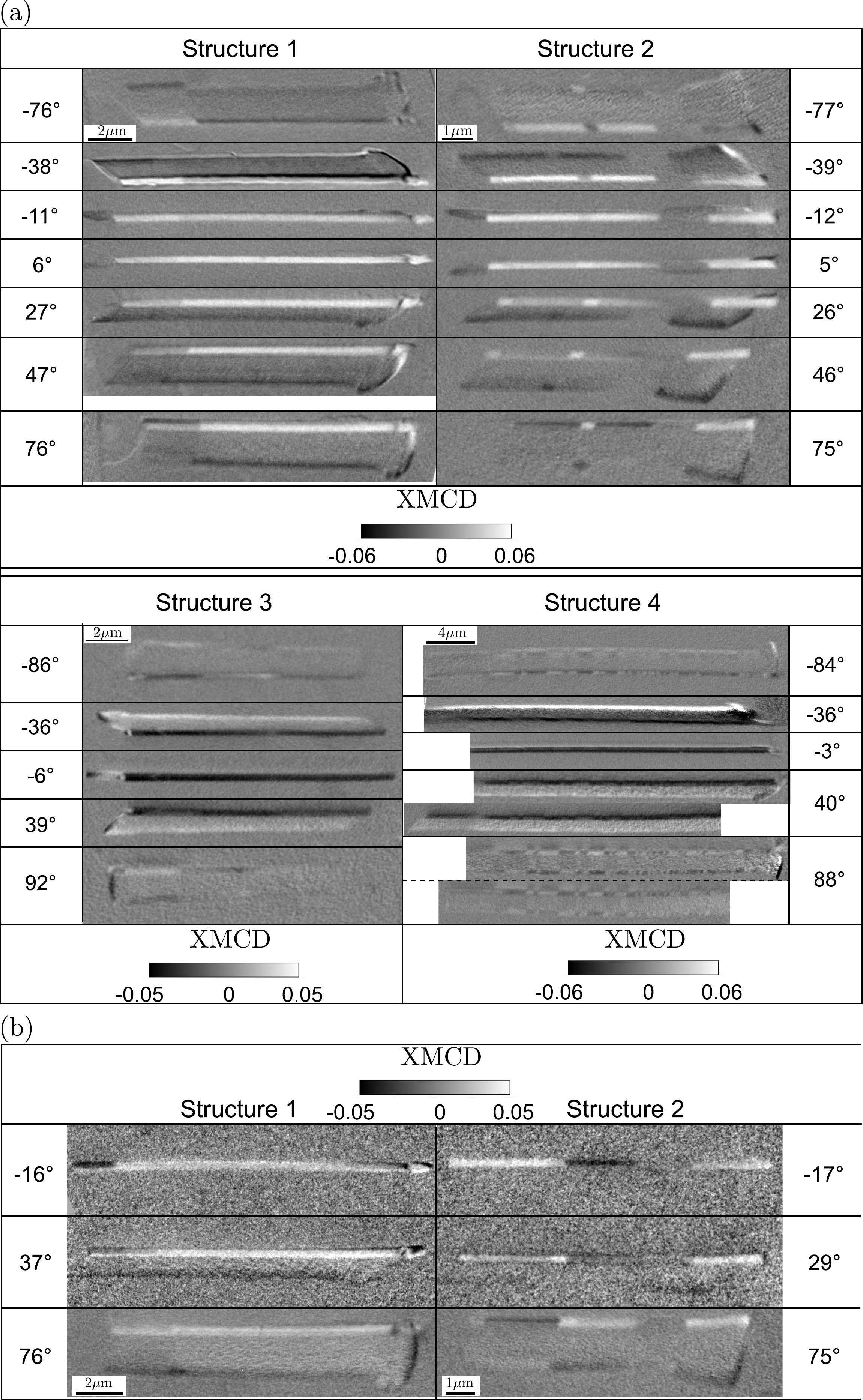}
\caption{
XMCD-PEEM images were recorded with different angles between the nanostructure's long axis and the X-ray beam. The images were used to plot the angular-dependent XMCD contrast in (a) Fig.1 and (b) Fig.2 in the main text. 
}
\label{Angular_XMCD-PEEM}
\end{figure*}

The XMCD-PEEM imaging was performed with several orientations between the X-ray beam and the long axis of the nanostructures, see Fig.~\ref{Angular_XMCD-PEEM}.

\begin{figure*}
\centering
\includegraphics[width=1\textwidth]{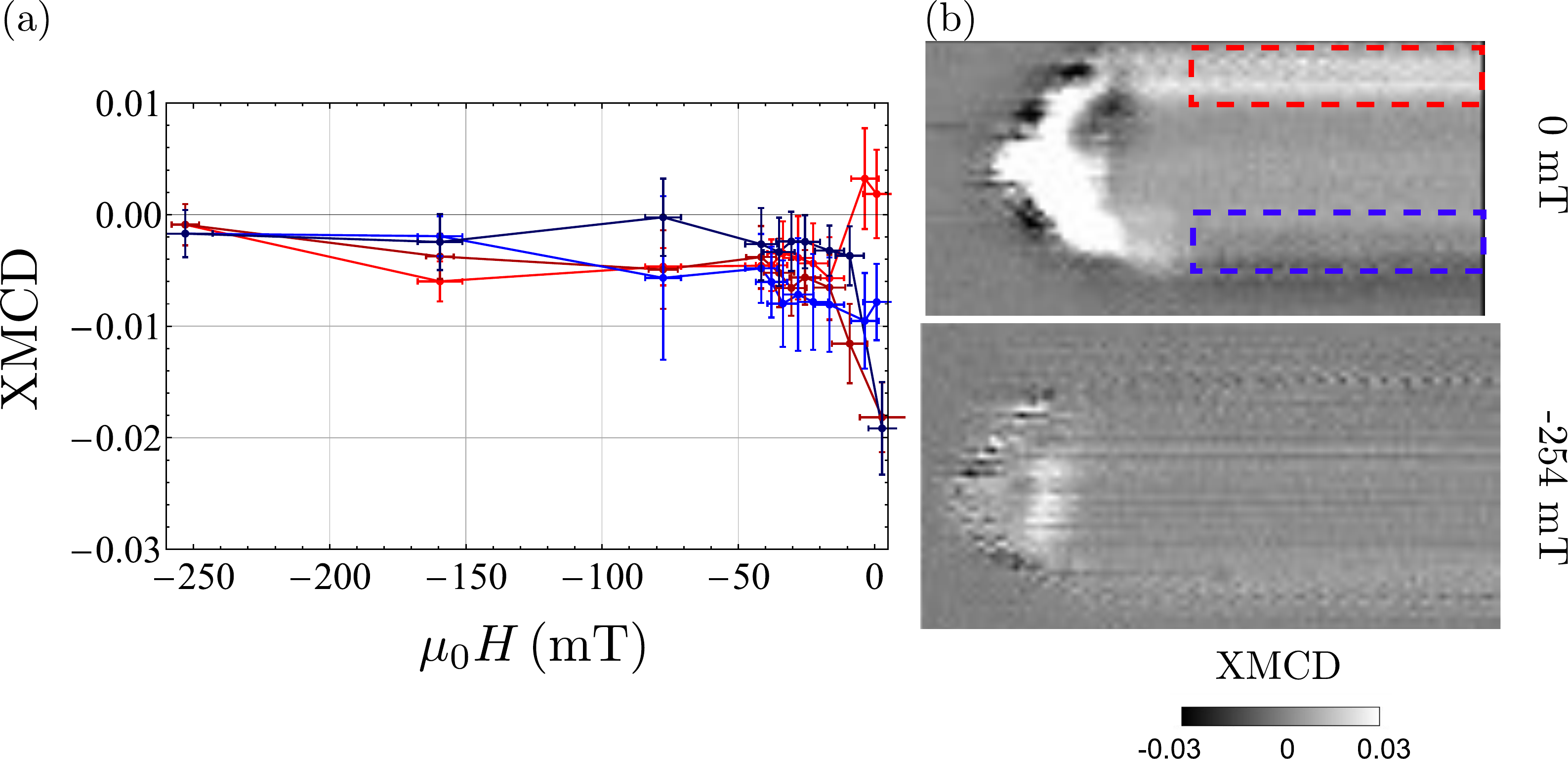}
\caption{
(a) The measured XMCD contrast from the nanostructure and (b) STXM images recorded at remanence before measuring a hysteresis loop and at a maximal applied field. The obtained XMCD contrast was averaged over the areas corresponding to the first pair of side facets in red and the opposite pair of side facets in blue.
}
\label{STXM-data}
\end{figure*}

Our investigation of the system’s anisotropies reveals a bi-axial anisotropy energy landscape with associated energy barriers that differ by an order of magnitude. Notably, the larger energy barrier corresponds to the magnetization switching under an applied magnetic field of approximately 10 mT. Therefore, we measure the hysteresis loop on a part of a single nanostructure utilising STXM-XMCD measurements, see Fig.~\ref{STXM-data}. for schematics. A $+ 250 $ mT magnetic field was applied before measurements to define a reference magnetic state. Then, STXM images were recorded under the application of a magnetic field up to $-253$ mT along the nanostructure`s long axis.
Assuming the chemical homogeneity of the FeCo alloy, a 100\% circularly polarised X-ray beam and the same intensity of the incoming beam for opposite polarizations $I_0(P = +1) = I_0(P=-1)$, one could write the connection between XMCD signal $I_\text{XMCD}$ and the averaged magnetization projection onto the beam propagation direction $\left<m_\text{proj}\right>$ as:
\begin{equation}
	\label{eq:STXM_contrast_to_magnetisation}
\left<m_\text{proj}\right> = \dfrac{\arctan \left( I_\text{XMCD}\right) }{t_\text{shell}\cdot \delta \mu_\text{FeCo}},
\end{equation}
with $t_\text{shell}$ the total length of the X-ray beam path through the ferromagnetic shell. The resonant magnetic contribution $\delta \mu_\text{FeCo}$ to the absorption coefficient for the $\text{Fe}\text{Co}$ alloy is calculated for at the Co $L_3$ absorption edge. The total length of the X-ray beam path through the magnetic film for the pairs of side facets is 25.6 nm. 
For the pairs of side facets, the longitudinal component of the magnetization is orthogonal to the X-ray beam and doesn't contribute to $\left<m_\text{proj}\right>$. Then, the contribution to $\left<m_\text{proj}\right>$ is done by the azimuthal component of the magnetization, assuming that the out-of-plane component is negligibly small.
One could use the model ${\sin \phi \cos\dfrac{\pi}{6} = \left<m_\text{proj}\right> }$, where the magnetisation angle $\phi$ lies in the facet's plane and is counted from the nanostructure's long axis. Then, the longitudinal component of the magnetization can be calculated as $	\left<	m_x\right>  = \cos\phi$.

To derive the law of approach to saturation for the GaAs-$\text{Fe}_{\text{33}}\text{Co}_{\text{67}}$ core-shell nanorods we follow the approach developed in Ref.~\cite{ZHANG20102375} and apply it using the anisotropy density energy landscape reported in Ref.~\cite{Muermann2008}.
Then, one could write the magnetisation component along the field direction as follows:
\begin{equation}
	\begin{array}{ccl}
	\label{eq:Law of approach}
  m_x = \cos \delta \approx 1 - \dfrac{\delta^2}{2},\\[2mm]

	\delta  \!=   \! \! \dfrac{-2\sqrt{2}\left(  K_1  \!+  \!4 K_U \right)\cos \!\left(2\Phi \right)  \!+  \! 2\sqrt{2} K_1 \cos \!\left(4\Phi \right) \! + \! \left[K_1 \! + \! 4 K_U  \!+ \! 7 K_1 \cos \!\left(2\Phi \right) \right] \sin \! \left(2\Phi \right) }{ 2 \!\left[ \! -6 H \mu_0 M_\mathrm{s}  \!+ \! \left(K_1 \!+ \! 4K_U \right) \!\cos \!\left(2\Phi \right)  \!+ \!K_1  \! \!\left[ 7  \cos \!\left(4 \Phi \right) \!- \! 4\sqrt{2} \sin \!\left(4\Phi \right)\right]   \! \! + \! 2\sqrt{2} \sin \!\left(2\Phi \right) \! \!\left(K_1 \! + \! 4 K_U \right)  \!    \right] },
	\end{array}
\end{equation}
where $H$ is the amplitude of the applied magnetic field, $M_\mathrm{s}$ is the saturation magnetization and $\Phi$ is an in-plane angle between the applied magnetic field and the nanostructure's long axis.

\bibliography{biblio}